# Development and validation of a cryogenic far-infrared diffraction grating spectrometer used to post-disperse the output from a Fourier transform spectrometer

Alicia M. Anderson  ; David A. Naylor  ; Brad G. Gom  ; Matthew A. Buchan  ; Adam J. Christiansen  ; Ian T. Veenendaal

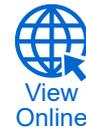
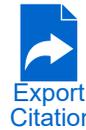

CrossMark

View Online

Export Citation

# Development and validation of a cryogenic far-infrared diffraction grating spectrometer used to post-disperse the output from a Fourier transform spectrometer

Alicia M. Anderson,[1,2,3,a)] David A. Naylor,[1] Brad G. Gom,[1] Matthew A. Buchan,[1,2] Adam J. Christiansen,[1,2] and Ian T. Veenendaal[4]

AFFILIATIONS
[1] Department of Physics and Astronomy, University of Lethbridge, 4401 University Dr. W, Lethbridge, Alberta T1K 3M4, Canada
[2] Blue Sky Spectroscopy Inc., Lethbridge, Alberta T1J 0N9, Canada
[3] Cavendish Laboratory, Astrophysics Group, University of Cambridge, J. J. Thomson Avenue, Cambridge CB3 0HE, United Kingdom
[4] School of Physics and Astronomy, Cardiff University, Cardiff CF24 3AA, Wales, United Kingdom

a)Author to whom correspondence should be addressed: aa2355@cam.ac.uk

ABSTRACT

Recent advances in far-infrared detector technology have led to increases in raw sensitivity of more than an order of magnitude over previous state-of-the-art detectors. With such sensitivity, photon noise becomes the dominant noise component, even when using cryogenically cooled optics, unless a method of restricting the spectral bandpass is employed. The leading instrument concept features reflecting diffraction gratings, which post-disperse the light that has been modulated by a polarizing Fourier transform spectrometer (FTS) onto a detector array, thereby reducing the photon noise on each detector. This paper discusses the development of a cryogenic (4 K) diffraction grating spectrometer that operates over the wavelength range of 285 to 500 $\mu$m and was used to post-disperse the output from a room-temperature polarizing FTS. Measurements of the grating spectral response and diffraction efficiency are presented as a function of both wavelength and polarization to characterize the instrumental performance.

## I. INTRODUCTION

As radiation travels through the universe, it experiences losses due to absorption and scattering from dust grains. Since the grain sizes are <1 $\mu$m,[1] observations at far-infrared (FIR) wavelengths (>30 $\mu$m) experience minimal losses and, thus, can probe obscured regions more efficiently. Spaceborne FIR spectroscopic observations provide a unique means of addressing some of the leading questions in modern astrophysics, from the formation of stars and planets in molecular clouds in our own galaxy[2] to the evolution of galaxies over cosmic time.[3]

The Fourier transform spectrometer (FTS), with broad spectral coverage (the multiplex advantage[4]), high throughput, and high and variable resolution, has been the instrument of choice for astronomical spectroscopy in the FIR (AKARI,[5] Herschel[6]). The exquisite sensitivity of modern superconducting detectors, having noise equivalent powers (NEPs) of $\sim 10^{-19}$ W/$\sqrt{\text{Hz}}$ (an improvement of over two orders of magnitude from those used on Herschel) necessitates that the next generation of spaceborne FIR observatories employ hybrid spectrometers. With photon noise now the dominant driver for spectrometer design, the previous multiplex advantage of FTS becomes a disadvantage unless the spectral bandwidth is restricted.

The solution adopted for the SPICA SAFARI[7] and PRIMA[8] instruments is to post-disperse the output from a FTS using diffraction grating spectrometers. Previously, grating spectrometers have been used alone to provide low-resolution broad-band spectroscopy in the FIR.[9,10]

In order to achieve the high dispersion required for the post-dispersed polarizing Fourier transform spectrometer (PDPFTS) concept, equivalent to a resolving power of $R \sim 300$, each diffraction grating must operate at high angles of incidence. At such angles, the grating response exhibits a strong polarization dependence, having high and uniform efficiency (~80%) for TM polarized light but lower and variable efficiency (10–40%) for TE polarized light. A FTS based on the Martin–Puplett polarizing interferometer[11] uses the polarizing encoding properties of the interferometer to optimally couple to the TM grating mode. When the FTS is scanned over its full optical displacement, each of the detectors measures a high resolution interferogram convolved with the grating spectral response function (SRF) for that particular detector/grating combination. Upon Fourier transformation, an individual interferogram yields a small bandwidth, high resolution spectrum. By stitching together the spectra from individual detectors, one FTS scan produces a high-resolution spectrum of the entire wavelength range. We refer to this system as a post-dispersed polarizing Fourier transform spectrometer (PDPFTS).[12] In this paper, we describe the design and performance of a cryogenic grating spectrometer that has been developed to explore the challenges of the PDPFTS technique.

## II. GRATING THEORY

The diffraction grating concept was first described in 1786 by Hopkinson and Rittenhouse,[13] who observed diffraction through a series of parallel wires. Fraunhofer extended this principle and ruled grooves onto an optical glass, which he used to study the solar spectrum.[14] The ruling process was completely transformed by Rowland, who developed several ruling engines and was able to create grating structures with resolving powers of ~150 000.[15,16]

With reference to Fig. 1, when monochromatic light of wavelength $\lambda$ is incident on a diffraction grating, it is diffracted into a discrete angle given by the grating equation,[17]

$$m\lambda = d(\sin \alpha + \sin \beta), \quad (1)$$

where $m$ is the order of diffraction. The right side of the equation represents the path difference between the light reflecting from adjacent grooves of the grating.[18] $\alpha$ and $\beta$ are the angles of the incident and diffracted light, measured with respect to the grating normal, and $d$ is the spacing between adjacent grooves.

The mounting configuration chosen for the grating spectrometer is the Czerny–Turner monochromator, as displayed in Fig. 1. In this configuration, the grating equation can be rewritten as

$$m\lambda = 2d(\sin(\theta - \phi)\cos\phi), \quad (2)$$

where $\phi$ is the deviation angle, and $\theta$ is the angle of incidence minus the deviation angle.[19]

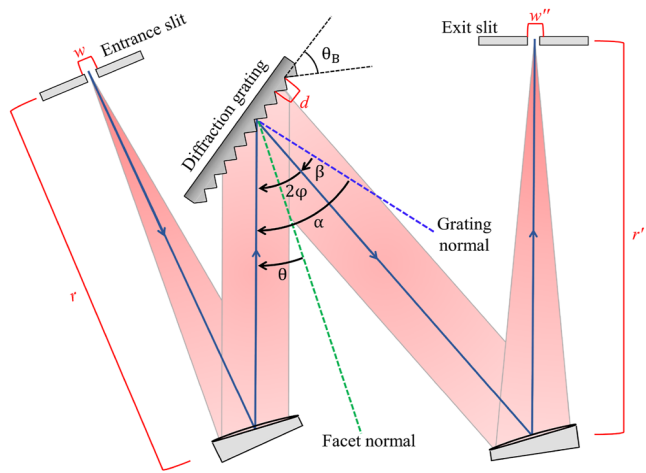

**FIG. 1.** Czerny–Turner monochromator configuration. $\alpha$ is the incident angle, $\beta$ is the diffracted angle, and the input and output collimating mirrors have focal lengths $r$ and $r'$, respectively. $w$ and $w''$ are the widths of the entrance and exit slits.

### A. Resolving power

For a generic spectrometer, the spectral resolving power is given by

$$R = \frac{\lambda}{\Delta\lambda}, \quad (3)$$

where $\Delta\lambda$ is the spectral resolution, often defined by the Rayleigh criterion.[17]

At a given wavelength, the resolution of a grating spectrometer is determined by the widths of the entrance and exit slits, as discussed below.

A linespread calculation is used to determine the convolution of the image of the entrance slit with the exit slit. As light passes through an entrance slit, it will form an image of the slit that will be magnified differently in the horizontal and vertical directions, a phenomenon referred to as anamorphic magnification.

In the dispersion direction, the tangential magnification of the width of the entrance slit is given by

$$\chi_T = \frac{w'}{w} = \frac{r'}{r}\frac{\cos\alpha}{\cos\beta}, \quad (4)$$

where $w$ is the width of the entrance slit, $w'$ is the width of the image of the entrance slit, and $r$ and $r'$ are the focal lengths of the entrance and exit optics, respectively.[18] To optimize the spectral resolution, the width of the exit slit, $w''$, should be chosen to match $w'$. When $w''$ is less than $w'$, the limiting resolution of the grating is achieved, but at a cost of lower throughput as the exit slit blocks some of the light from reaching the detector. Contrarily, when $w''$ is greater than $w'$, the exit slit width limits the resolution.

The resolution of a grating spectrometer limited by the slit widths can be expressed as[18]

$$\Delta\lambda = \frac{\max(w'', w')d\cos(\theta - \phi)}{mr'}, \quad (5)$$

where $\max(w'', w')$ denotes the maximum between the exit slit, $w''$, and $w'$. For the grating designed in this study, $w''$ was chosen to match $w'$ at the center of the band, 392.5 $\mu$m. The theoretical resolving power falls into two regimes:

$$R = \begin{cases} \dfrac{\lambda m r}{dw \cos(\theta + \phi)} & \text{for } w' < w'' (\lambda < 392.5\,\mu\text{m}) \\ \dfrac{\lambda m r'}{dw' \cos(\theta - \phi)} & \text{for } w' > w'' (\lambda > 392.5\,\mu\text{m}) \end{cases}. \quad (6)$$

In the top case, the system is entrance slit limited, and in the bottom case, it is exit slit limited. This analysis assumes that the system can be understood in terms of geometrical optics, which is a fair approximation since the components are oversized with respect to the wavelength of light ($\lambda/d \gtrsim 10$). Equation (6) has been used to model the theoretical resolving power of the grating described in this paper.

### B. Efficiency

When observing faint astronomical sources, it is important to maximize the energy diffracted into the order of interest. This can be accomplished by a process known as blazing, where the grooves of a grating are shaped to maximize efficiency at a particular wavelength. The blaze angle is given by[18]

$$m\lambda_B = 2d \sin \theta_B, \quad (7)$$

where $\theta_B$ is the blaze angle labeled in Fig. 1. $\lambda_B$ is the wavelength where the efficiency is maximized, and in our case, it is chosen to be the band center at 392.5 $\mu$m.

Theoretical modeling of the grating efficiency requires knowledge of the electromagnetic fields exterior to, within, and at the boundary surface of the grating substrate. In principle, these fields can be determined by solving Maxwell's equations at the boundary. However, the diffraction formulas derived from Kirchhoff's approximations are not valid when the groove spacing is on the order of the wavelength of radiation,[17] as is the case for this study. Rigorous theoretical models have been developed over the past 70 years,[20–24] although solving Maxwell's equations for a boundary surface with an arbitrary profile is computationally challenging. In 1980, the problem was simplified by Chandezon et al.,[25] who described a method that applies a translation of the coordinate system to map the grating profile to a planar surface, greatly simplifying the boundary conditions. This technique exploits the periodic nature of the profile to implement Fourier analysis in solving Maxwell's equations, reducing the calculation to an eigenvalue problem.[25] The methods discussed by Li et al.[26] were used to produce a theoretical polarization-sensitive efficiency model of the diffraction grating presented in this paper.

### III. DESIGN OF THE GRATING MODULE

The grating for the PDPFTS was designed to operate over a wavelength range from 285 to 500 $\mu$m, chosen to match available test equipment. The theoretical resolving power, $R$, was calculated using Eq. (6) under the assumption that the spectrometer was slit width limited. The width of the exit slit, $w''$, was chosen to achieve a resolving power of $R \sim 100$ at the middle of the wavelength range (392.5 $\mu$m), and the grooves were blazed to maximize the efficiency at this wavelength. Using Eq. (7), the corresponding blaze angle is 39.4°. The specifications of the grating, designed to operate in first order, are listed in Table I. The grating was fabricated from Rapidly Solidified Aluminum 6061 (RSA6061) and ruled with a single point diamond machine under a specialized thermalization process to minimize internal stress.

**TABLE I.** Specifications for the grating spectrometer designed to operate over the range of 285–500 $\mu$m.

| Parameter | Value |
|---|---|
| Grating dimensions (width × length) | 105 × 50 mm$^2$ |
| Slit spacing, $d$ | 312 $\mu$m |
| Blaze angle, $\theta_B$ | 39.4° |
| Deviation angle, $2\phi$ | 15° |
| Entrance slit width, $w$ | 5.0 mm |
| Exit slit width, $w''$ | 4.0 mm |
| Entrance focal length, $r$ | 310 mm |
| Exit focal length, $r'$ | 310 mm |

The design of the grating spectrometer builds on the work of Veenendaal et al.,[27,28] who developed a cryogenic post-dispersing grating to sort the output from a Fabry–Pérot interferometer. In the new design, the grating is mounted in a monolithic aluminum enclosure on a pivot driven by a cryogenic stepper motor through worm gear reduction. The monolithic design and material choice for the enclosure ensure that the system maintains alignment (e.g., gear drive and optics) as the system is cooled to cryogenic temperatures. As envisaged, the far-infrared post-dispersed polarizing FTS instrument concept will incorporate several stationary diffraction gratings to distribute the signal from a polarizing FTS across different spectral bands of interest onto an array of ultra-sensitive detectors. However, since we did not have access to a detector array, the grating needed to rotate to change the angle of incidence/diffraction and scan the wavelength range of interest using a single detector.

The monolithic enclosure that houses the grating is shown schematically in Fig. 2. Improvements from the previous design include a monolithic grating enclosure and shields (teal, 1), which have a highly reflective exterior to reduce absorption and minimize thermal loading as the system is cooled to 4 K. The inside surfaces of the enclosure were coated with epoxy and sprinkled with carborundum particles to reduce reflectivity, thereby mitigating stray light. We adopted the method that was developed to coat optical components at the Herschel Space Observatory.[29] Other notable improvements include a larger diameter low-pass filter (50 mm) (brown, 2), a new diffraction grating (yellow, 3) that features a plane mirror mounted to the rear side (red, 4) free to rotate 360°, and a retractable baffle (magenta, 5) to block stray light within the grating enclosure from reaching the detector. The new 50 mm windows have custom low-pass filters with a cut-off frequency of 35 cm$^{-1}$. An additional low-pass filter is mounted at the interface between the 4 and 0.3 K enclosures to further limit stray light contaminating the detector signal. The exit slit (orange, 6) is mounted on the feedhorn of a 0.3 K composite bolometer detector (purple, 7). The cryogenic stepper motor (blue, 8) drives the worm and gear system and rotates the grating around the axis indicated by the black arrow.

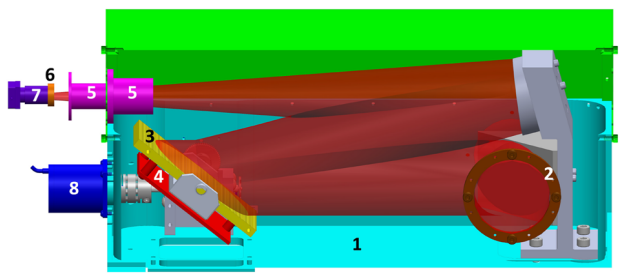

**FIG. 2.** Cut-away view of the grating spectrometer. See the text for details.

TABLE II. Specifications for the calibration FTS.

| Parameter | Value |
| --- | --- |
| Spectral range | 10–333 cm$^{-1}$ (1000–30 $\mu$m) |
| Spectral resolution | 0.016 cm$^{-1}$ |
| Maximum OPD | 32 cm |
| Resolving power | 625–20 813 |
| Throughput | 0.14 cm$^2$ sr |
| Vacuum pressure | <10 mTorr |

The grating assembly is clamped to the 4 K baseplate of a test-facility cryostat.[30,31]

### A. Rear mounted mirror

The derivation of the diffraction efficiency for a blazed reflection grating was discussed in Sec. II B. While these models provide a means to model the efficiency response, they do not account for manufacturing imperfections. There is literature available with efficiency measurements of various diffraction grating geometries;[32–34] however, the grating used in this study was custom-made to operate over far-infrared wavelengths and in a cryogenic environment. Thus, the efficiency could not have been characterized prior to receiving it because that would require an extensive suite of test equipment. We devised a simple method of determining the grating efficiency by mounting a mirror on the rear of the grating saddle. When the system is coupled to an FTS and the mirror is inserted into the optical path by rotating the grating 180°, a single measurement of the entire band is obtained, which serves to calibrate the efficiency of the grating as a function of wavelength.

### B. Source module

To produce realistic synthetic astronomical spectra, the source module should consist of a broad-band emission source (continuum) and a narrow-band line source (spectral feature). The source module used for the results presented in Secs. IV and V includes a commercial blackbody (300–1200 K) and two quasi-monochromatic line sources. The line sources are custom built and feature two THz photomixers,[35,36] each illuminated by two independently tunable, continuous-wave infrared lasers operating in the 1550 nm band. The beat frequency between the two lasers modulates the conductivity of the photomixer material, and the embedded antenna generates THz radiation, which is coupled to free space by a hyper hemispherical silicon lens. The lens sits adjacent to the photomixing module, allowing the focus to be placed 22.3 ± 1.2 mm behind the face of the lens, which facilitates the mounting of the device. The output from the lens is a cone of radiation with a wavelength-dependent 12–15° opening angle. However, at the wavelengths used in the results section, the device always has a 15° output angle. With the current suite of lasers, this optical heterodyne technique allows us to produce quasi-monochromatic lines in the region from 0.2 to 1.5 THz whose line widths are two orders of magnitude below the resolution of the PDPFTS—amply sufficient to explore its performance.

### C. Calibration FTS

A room temperature polarizing calibration FTS (cFTS), provided by Blue Sky Spectroscopy Inc.,[37] was used for the measurements presented in this paper. The optical configuration of the cFTS is a Martin–Puplett interferometer, MPI, and it was developed as a calibration facility to characterize THz sources (line or continuum), components (filters, samples, etc.), and detectors (single-pixel or arrays). Due to these applications, the spectrometer operates over a broad spectral range with high throughput and high spectral resolution power, as listed in Table II. In normal operating conditions, the chamber of the cFTS, including the path of the translating mirror up to the maximum optical path difference (OPD) of 32 cm, is evacuated to <10 mTorr, which prevents significant signal depletion due to molecular absorption from air. In addition, the input polarizer, polarizing beamsplitter, and output analyzer of the cFTS provide a combined efficiency of >95% over the entire spectral range.[38] A key advantage of coupling a polarizing FTS to a diffraction grating spectrometer is the ability to orient the output analyzer to match the more efficient TM mode of the grating.

## IV. RESULTS I: EFFICIENCY

Figure 3 shows the optical configuration for the results presented in this section of the paper. The dashed lines indicate the separation between room temperature (300 K) components, the cryogenic (4 K) grating assembly, and the composite bolometer detector chamber (300 mK). Low-pass optical filters[39] are mounted to the 100 and 4 K thermal shields and the grating and detector entrance apertures to reduce thermal loading and limit the bandwidth of radiation reaching the detector. Figure 4 shows a CAD schematic and image of the optical components mounted in the cryostat. The source was placed at the input of the FTS at room-temperature, and the output of the FTS was brought to a focus on the cryogenic entrance slit (1) using a custom $f/690°$ off-axis parabolic mirror (OAP). After passing through the slit, the beam was collimated by a $f/615°$ OAP (2), reflected by a flat mirror (3) toward a pendulum mirror (4), before passing through a 58 cm$^{-1}$ low-pass filter (5) and into the grating module. Light from the grating was then measured by the composite bolometer detector (6).

Measurements were obtained with the hybrid room-temperature/cryogenic PDPFTS described above. The source module, comprising the blackbody and tunable photomixer(s), operated at atmospheric pressure a short distance (∼ 8.6 cm) in front of the evacuated FTS. To study the instrument line shape

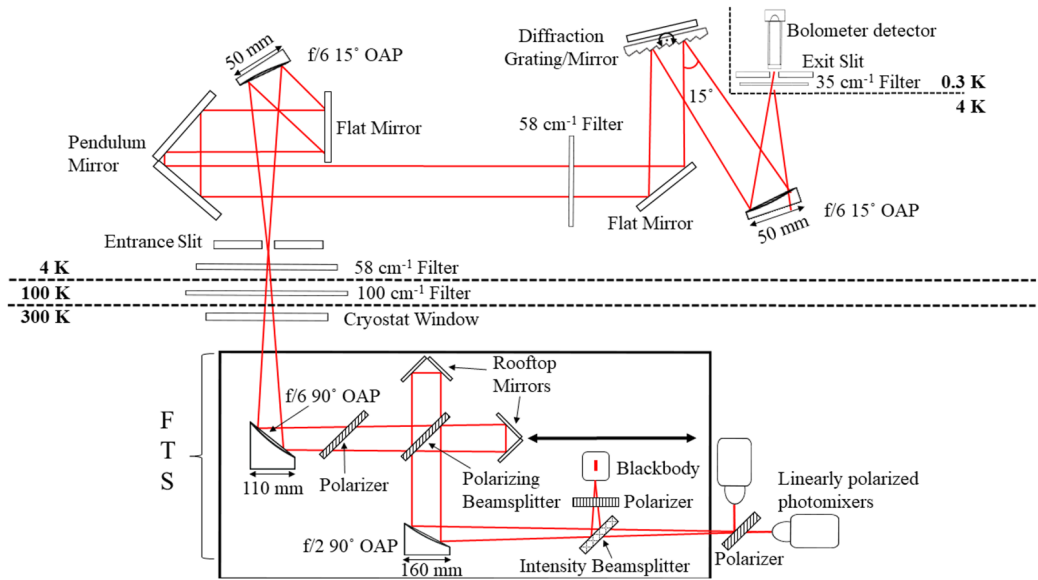

**FIG. 3.** Diagram of the optical setup used to measure the resolving power and efficiency of the diffraction grating spectrometer. The photomixer sources and external polarizer were placed a short distance (∼ 8.6 cm) outside of the evacuated FTS.

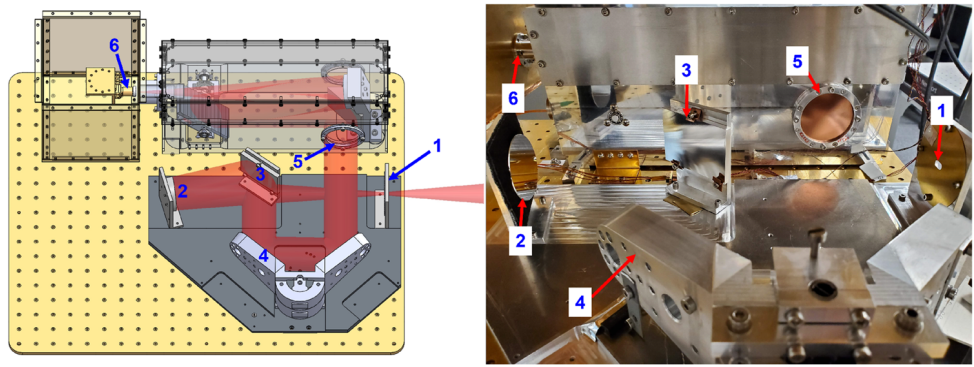

**FIG. 4.** CAD model (left) and an image (right) of the optical components within the 4 K chamber of the test facility cryostat. The numbered labels follow the light path. See the text for details.

(ILS) of the PDPFTS, both emission and absorption line measurements were obtained. By differing spectroscopic measurements of the blackbody and photomixer from those of the blackbody alone, it is possible to probe the ILS of the PDPFTS using the quasi-monochromatic photomixer sources. The left column of Fig. 5 shows the PDPFTS spectra of the blackbody source with the photomixer tuned to ∼32.2 cm$^{-1}$ (blue) in emission (bottom) and absorption (top). The photomixer was subsequently turned off while the spectra of the blackbody alone were measured (red). The difference between these measurements is shown as the green points in the right panel, which have been fitted with the sinc function (blue). Table III presents the FWHM and center frequency extracted from the sinc fits to both the absorption and emission lines.

**TABLE III.** FWHM and center frequency extracted from the sinc fits to the emission and absorption photomixer lines in Fig. 5. The fitted values are compared with the theoretical resolution of the FTS as determined by the maximum OPD.

|  | FWHM (cm$^{-1}$) | Center frequency (cm$^{-1}$) |
|---|---|---|
| Absorption | 0.016 05 ± 0.0011 | 32.174 ± 0.006 |
| Emission | 0.016 05 ± 0.0011 | 32.175 ± 0.006 |
| Theoretical | 0.016 04 | N/A |

A unique attribute of coupling a polarizing FTS to a diffraction grating spectrometer is the ability to measure the efficiency of the grating as a function of both wavelength and polarization. For these measurements, a calibrated cavity blackbody source with a slit geometry matched to the entrance slit of the grating was placed at the input to the FTS. When the mirror on the rear side of the grating was placed in the optical path of the incident beam, a scan of the FTS was used to obtain a measurement of the entire spectrum. The grating was then rotated into the path, and the incident angle, $\theta$, was varied to change the wavelength being measured by the detector. At each grating angle, the FTS was scanned five times, and the interferograms were averaged, phase-corrected, and Fourier transformed to yield spectra at each grating position.

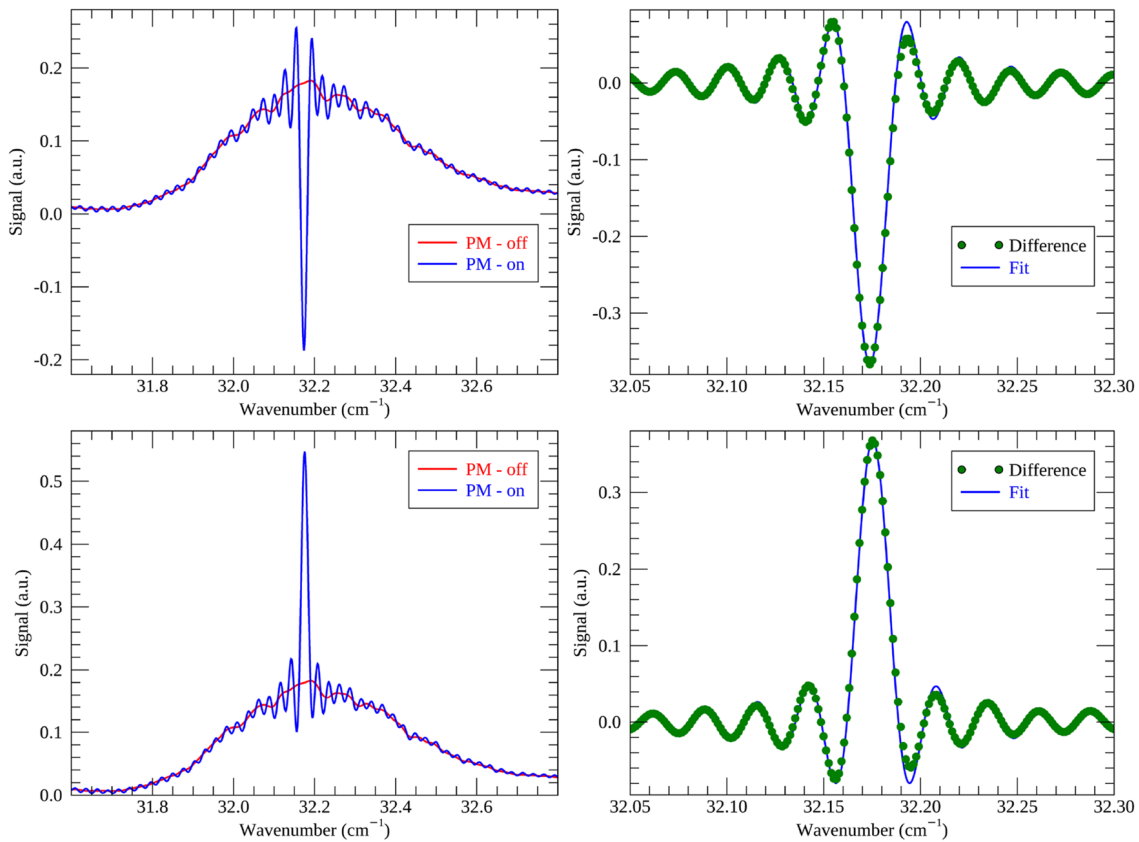

**FIG. 5.** Left column: spectra of blackbody and line emission/absorption (blue) and blackbody alone (red). Right column: difference spectra (green points) and fit to the FTS ILS sinc function (blue).

Fulton et al.[40] discuss the challenges of phase correcting post-dispersed data, which contain only a few points—roughly ten or fewer, depending on the spectral bandwidth presented by the diffraction grating and the resolution of the FTS. Since the instrumental phase of the FTS is common to all grating measurements across the band, they can be combined to determine the phase correction function more robustly.[40] The phase-corrected grating spectra and corresponding mirror spectra are shown in Fig. 6. The top panel shows FTS spectra polarized perpendicular to the grating grooves, i.e., in the dispersion direction (s-polarized). The bottom panel shows FTS spectra polarized parallel to the grating grooves (p-polarized). The varying background in the mirrored spectrum (black curve) in each plot is due to molecular absorption within the ~ 8.6 cm of the atmospheric path and channel fringes caused by the vacuum chamber windows. A closer inspection in Fig. 6 shows that these channel fringes are also present in the individual grating scans.

When the flat mirror is inserted into the optical path, radiation from across the entire spectral band falls onto the detector. By comparison, when the grating is in the optical path, the bandwidth of the signal is significantly reduced (i.e., less than 1% of the power received with the mirror in position). Similar to all bolometer detectors, the responsivity is a function of radiant loading and will be different when viewing the entire spectral band compared to viewing a narrow spectral region. The nonlinear response of the bolometer has been well studied under various conditions of radiant power loading.[41] By measuring the bolometer voltage and bias current while loaded with the wide spectral range and the narrow range, estimates of the radiant loading in each case can be inferred using the known performance of the detector. This allowed us to apply a first-order correction to the measured spectra to account for the nonlinear response. The corrected data are shown in Fig. 6, from which the grating efficiency, as a function of both wavelength and polarization, could be determined.

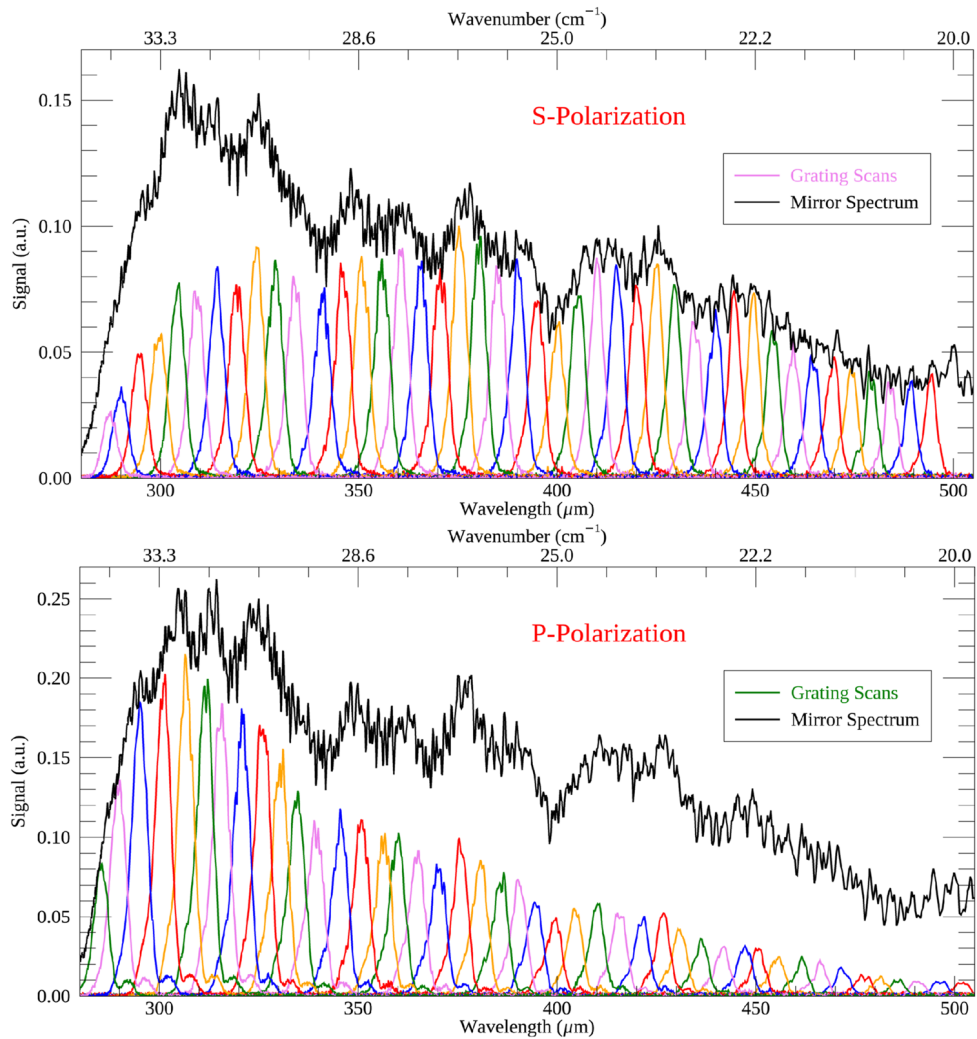

**FIG. 6.** FTS scans (colored lines) measured with the blackbody source at different angular positions of the grating. An FTS measurement of the entire band (black) was taken with the flat mirror in place of the grating. Features present in the mirror spectrum (black) are due to atmospheric absorption and channel fringes arising from the vacuum chamber windows.

Figure 7 presents the efficiency measurements calculated for each polarization state by comparing the amplitude of the grating scan to the signal amplitude of the mirror data at the same wavelength position. The data were corrected for the nonlinear response of the detector and multiplied by a scaling factor, which comprises all unknown efficiency losses between the grating and mirror systems. Thus, the measurements shown in Fig. 7 were taken to probe the general trend of the grating efficiency; they are not a measurement of the absolute efficiency, which would require a more detailed analysis of the potential coupling inefficiencies and a more robust determination of the detector response factor. The measurements are compared with the theoretical model for both polarization states.[25]

The s-polarization (TM mode) diffraction efficiency is shown to be greater than the p-polarization (TE mode) efficiency for a significant portion of the grating band, where $\lambda > 330$ $\mu$m. We were able to reproduce the models well, although we suspect the deviations from the theoretical model at short wavelengths are likely due to machining imperfections (periodicity, groove spacing), which have a more significant impact on the theoretical curves when the wavelength approaches the groove spacing ($d = 312$ $\mu$m). To our knowledge, these represent the first measurements reporting the diffraction efficiency of a blazed grating as a function of wavelength and polarization at cryogenic temperatures and far-infrared wavelengths.

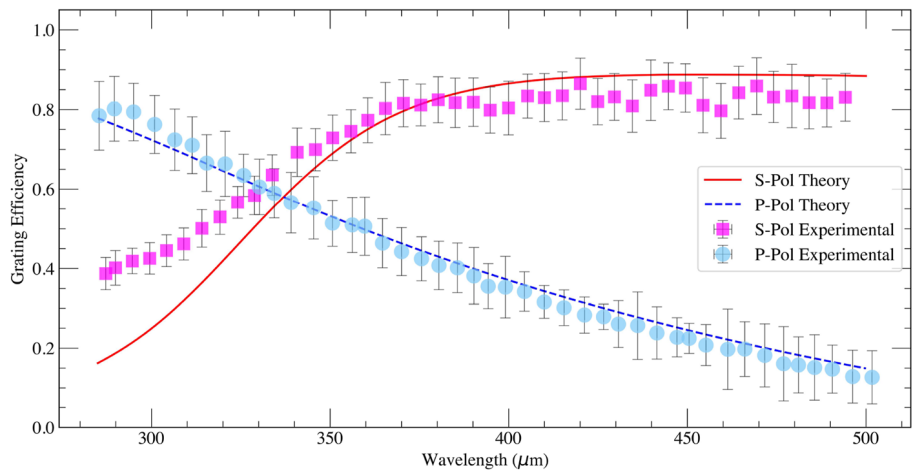

**FIG. 7.** Diffraction efficiency measurements of s-polarized (pink hexagons) and p-polarized (blue circles) light. The measured data are compared to numerical calculations of the diffraction efficiency for s-polarized (red curve) and p-polarized (blue dash) light using the coordinate-transform method.

**TABLE IV.** Specifications for the fully cryogenic grating spectrometer measurements. The parameters in bold are changed from the previous setup.

| Parameter | Value |
| --- | --- |
| Grating dimensions (width × length) | $105 \times 50$ mm$^2$ |
| Slit spacing, $d$ | 312 $\mu$m |
| Blaze angle, $\theta_B$ | 39.4° |
| Deviation angle, $2\phi$ | 15° |
| **Entrance slit width**, $w$ | **3.0 mm** |
| Exit slit width, $w''$ | 4.0 mm |
| **Entrance focal length**, $r$ | **190.6 mm** |
| Exit focal length, $r'$ | 310 mm |

## V. RESULTS II: RESOLVING POWER

Ultimately, the goal of this work is to implement a fully cryogenic PDPFTS instrument. Building on the results from the previous section, the final experimental configuration presented in this paper sought to combine a cryogenic line source with the grating and bolometer detector. This configuration serves to deploy three out of the four instrument modules at cryogenic temperatures. In this configuration, we are able to evaluate the performance of the diffraction grating spectrometer and the line source in a fully cryogenic environment. The performance metric we explored was the variation in resolving power as a function of wavelength.

The specifications of the cryogenic testing configuration are listed in Table IV. The differences between this design and the previous one (see Table I) are displayed with bold text. Figure 8 shows a schematic of the optical design for the results presented in this section. A tunable line source was provided by a low-temperature grown gallium arsenide (LTG GaAs) terahertz photomixer.[42] Figure 9 shows a CAD rendering (top) and image (bottom) of the optical components in the cryostat. The photomixer (1) outputs a collimated beam with a diameter of ∼ 8 mm. The collimated beam is focused by a 25.4 mm $f/1$ 90° OAP (2) through the entrance slit (3) and collimated by a 50.8 mm $f/3.75$ 90° OAP (4). The collimated beam passes through a 58 cm$^{-1}$ low-pass filter (5) mounted to the entrance aperture of the grating enclosure (6),

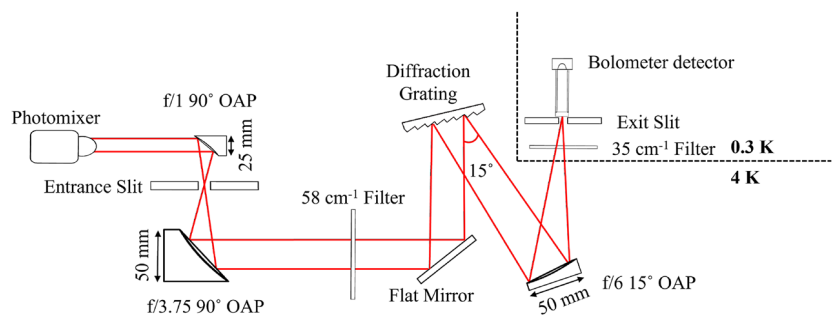

**FIG. 8.** Diagram showing the optical setup used to measure the resolving power and efficiency of the diffraction grating spectrometer with a cryogenic line source.

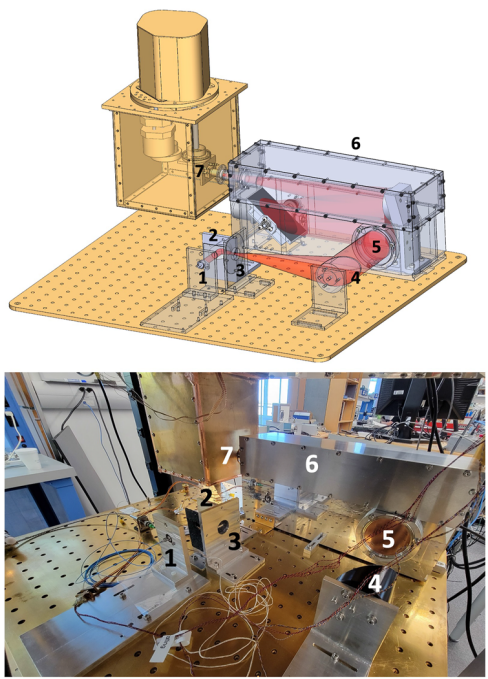

**FIG. 9.** CAD model (top) and an image (bottom) of the optical components for the fully cryogenic testing configuration. The numbered labels follow the photon path of the light. See the text for details.

and the grating disperses the light onto the exit slit located on the bolometer feedhorn (7).

The photomixer used in these measurements employed two 780 nm continuous-wave laser diodes[43] operating at slightly different frequencies. An AC photocurrent is induced in the photomixer at the optical beat frequency when a bias voltage is applied. By modulating the bias voltage with an arbitrary waveform generator, an amplitude modulated optical signal from the photomixer is produced. A state-of-the-art low-noise differential pre-amplifier described in Ref. 41 measures the detected signal from opposite sides of the symmetric bolometer element in a fully differential configuration to eliminate common-mode noise. The differential AC signal was measured by a model SR830 digital signal processing (DSP) lock-in amplifier (LIA) with the reference signal provided by the waveform generator. The output from the LIA was digitized and recorded as a function of the grating position.

For the measurements presented in this section, the photomixer was tuned across the wavelength range from 285 to 479 $\mu$m. At each setting of the photomixer, the grating was scanned in 0.06° increments (0.58–0.40 $\mu$m) around the corresponding photomixer wavelength to determine the spectral response function (SRF) of the grating. The far-infrared wavelength emitted by the photomixer was determined using a wavemeter[44] to measure the wavelengths of the individual 780 nm lasers, whose difference corresponded to the expected output wavelength. The normalized grating SRFs are shown in Fig. 10. Each SRF was fitted with a Gaussian function,

$$f(\lambda) = A_0 \exp\left(\frac{-(\lambda - \lambda_c)^2}{\Delta\lambda^2}\right) + A_1 \quad (8)$$

to determine the center wavelength, $\lambda_c$, and full-width-half-maximum, $\Delta\lambda$. The experimental resolving power, $R$, was determined using Eq. (3).

Figure 11 shows data from three grating SRFs along with the best-fit Gaussian profile for each dataset. From left to right, the data shown were collected at photomixer wavelengths of 304, 394, and 474 $\mu$m. It is evident from Fig. 11 that there are extraneous contributions in the wings of the profiles, which we attribute to stray reflections reaching the detector. To illustrate this effect, Fig. 12 shows all grating SRFs transposed to the same angular scale and over-plotted around 0°. Features outside of the expected grating profile appear at approximately the same angular positions (−1°, −0.7°, and 0.6°), irrespective of the different photomixer wavelengths. These features have been traced to stray reflections from mounting components within the spectrometer that will be mitigated by applying absorbing coatings to the components and do not significantly impede our ability to recover the resolving power as a function of wavelength. Figure 13 shows the resolving power measured from all grating SRFs. The data are compared with the theoretical slit-width limited resolving power in Eq. (6). The data follow the overall trend of the theoretical curve. Even with a slightly lower resolving power, these measurements show that the grating succeeds as a post-dispersing module by restricting the spectral band of radiation viewed by the detector, achieving the target $R \sim 100$ at the band center.

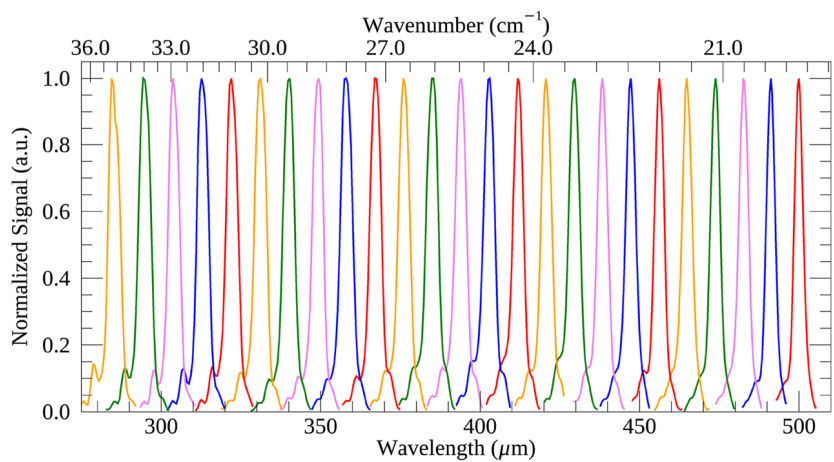

**FIG. 10.** Normalized grating SRFs taken with the photomixer source. The data were measured with the output polarization of the FTS configured perpendicular to the grating grooves (s-polarized). The profiles show a slight narrowing with increasing wavelength, as expected.

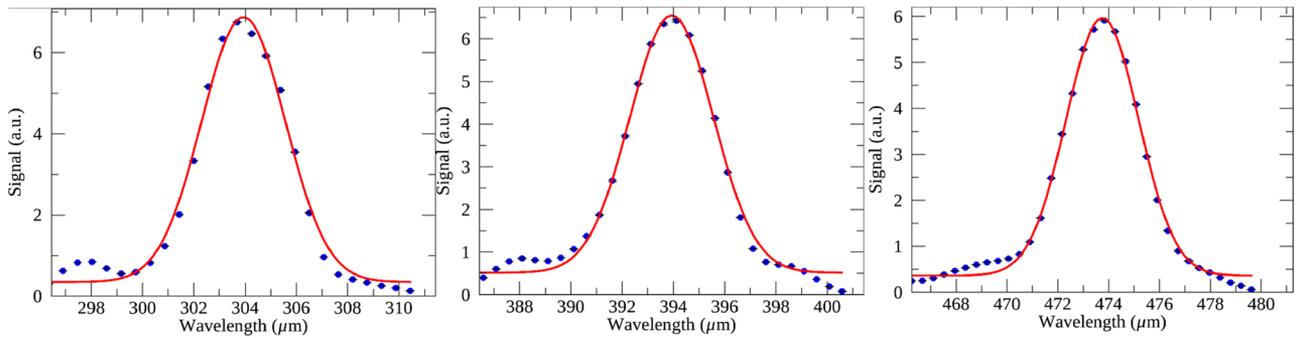

**FIG. 11.** Measured data (blue points) and the best-fit Gaussian profile (red curve) for grating SRFs measured with the cryogenic line source.

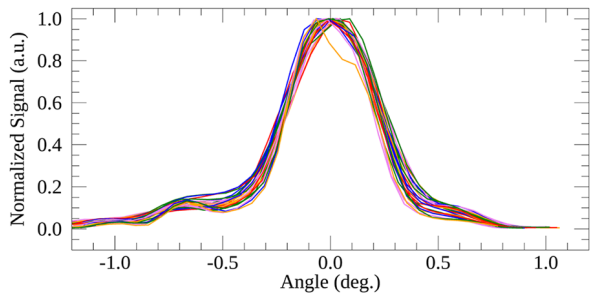

**FIG. 12.** All grating SRFs over-plotted in an angular position and centered at 0°.

## VI. CONCLUSIONS AND FUTURE WORK

The design and performance of a cryogenic far-infrared grating spectrometer have been presented. We have described a novel method to measure the grating efficiency curve as a function of both polarization state and wavelength at cryogenic temperatures and far-infrared wavelengths using a polarizing FTS. The results show that the diffraction grating spectrometer produced a spectral response that agrees well with the theory.

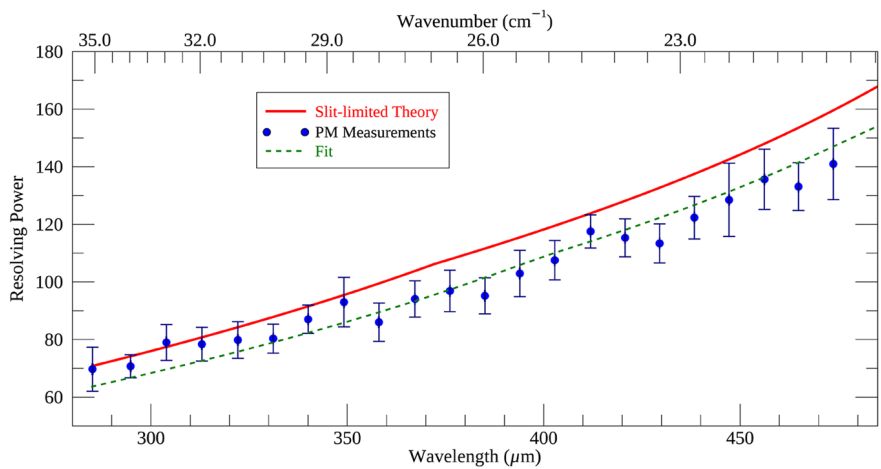

**FIG. 13.** Experimental resolving power of the grating measured with the cryogenic photomixer (blue circles). The theoretical resolving power (red) is described by Eq. (6), and the data were fit (green) with the same equation to extract a multiplication factor. The fit returned a factor of 0.92 compared to the theory.

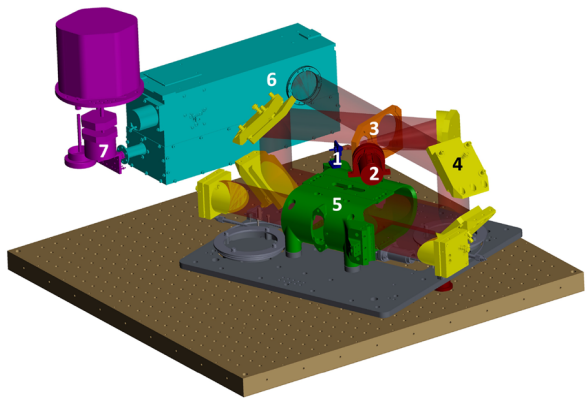

**FIG. 14.** CAD rendering of the cryogenic PDPFTS within the 4 K volume of a large facility cryostat.

We have independently verified the spectral resolving power of the diffraction grating in a fully cryogenic (4 K) environment with the second set of measurements. These results have shown the first successful integration of a cryogenic line source with our diffraction grating spectrometer and bolometer detector.

While the results presented in this paper demonstrate the performance of the grating spectrometer, the final PDPFTS configuration will employ a cryogenic polarizing FTS to couple with the cryogenic source, grating, and bolometer. The fully cryogenic configuration benefits by eliminating chamber windows and, thus, channel fringes, as shown in the data presented in Fig. 6. Figure 14 shows a schematic of the cryogenic PDPFTS that is currently under development. The scanning FTS mechanism (FTSM) is provided by ABB Inc.,[45] and the source module is comprised of the cryogenic photomixer described in this paper (blue, 1) coupled to a cryogenic blackbody source (red, 2) by a linear polarizer (orange, 3). Auxiliary optics (yellow, 4) couple light from the source into the FTSM (green, 5) and then into the grating spectrometer (teal, 6), which directs light toward the bolometer detector (purple, 7). The culmination of this project will be achieved upon the successful integration of the cryogenic source module and FTSM to realize the first fully cryogenic PDPFTS.

## ACKNOWLEDGMENTS

This research was funded in part by Alberta Innovates, Blue Sky Spectroscopy Inc., CFI, CMC, the CSA, NSERC, NTCO, and the University of Lethbridge. The authors would like to thank members of the Institute for Space Imaging Science at the University of Lethbridge and Blue Sky Spectroscopy, Inc.

## AUTHOR DECLARATIONS
### Conflict of Interest

The authors have no conflicts to disclose.

### Author Contributions

**Alicia M. Anderson**: Data curation (equal); Formal analysis (equal); Visualization (equal); Writing – original draft (equal). **David A. Naylor**: Conceptualization (equal); Funding acquisition (equal); Project administration (equal); Resources (equal); Supervision (equal); Writing – review & editing (equal). **Brad G. Gom**: Conceptualization (equal); Project administration (equal); Resources (equal); Supervision (equal); Writing – review & editing (equal). **Matthew A. Buchan**: Data curation (supporting); Investigation (supporting); Visualization (supporting); Writing – review & editing (equal). **Adam J. Christiansen**: Writing – review & editing (equal). **Ian T. Veenendaal**: Conceptualization (equal); Resources (equal).

## DATA AVAILABILITY

The data that support the findings of this study are available from the corresponding author, A. M. Anderson, upon reasonable request.